\documentclass[twocolumn,prb,showpacs]{revtex4}

\usepackage{graphicx} \usepackage{rotating} \usepackage{amssymb}

\newcommand{\la}{\mathord{\langle}}
\newcommand{\ra}{\mathord{\rangle}}

\begin{document}

\title{In-plane dipole  coupling anisotropy of  a square ferromagnetic
Heisenberg monolayer}

\author{M.  Dantziger}  \author{B.  Glinsmann}  \author{S.  Scheffler}
\author{B. Zimmermann} \author{P.  J. Jensen} 
\altaffiliation[On leave from the ]{Institut f\"ur Theoretische Physik, 
Freie Universit\"at Berlin, Arnimallee 14, D-14195 Berlin, Germany} 
\email[\\ Corresponding author.  Electronic adress: ]
{jensen@physik.fu-berlin.de}

\affiliation{I.\  Institut  f\"ur  Theoretische Physik,  Universit\"at
Hamburg, \\ Jungiusstr.9, D-20355 Hamburg, Germany}

\begin{abstract} 
In  this  study   we  calculate  the  dipole-coupling-induced  quartic
in-plane  anisotropy of a  square ferromagnetic  Heisenberg monolayer.
This anisotropy increases with an increasing temperature, reaching its
maximum value close to the  Curie temperature of the system.  At $T=0$
the system is  isotropic, besides a small remaining  anisotropy due to
the zero-point motion of quantum mechanical spins.  The reason for the
dipole-coupling-induced  anisotropy is the  disturbance of  the square
spin   lattice  due   to  thermal   fluctuations  ('order-by-disorder'
effect). For usual  ferromagnets its strength is small  as compared to
other anisotropic  contributions, and  decreases by application  of an
external magnetic  field. The results  are obtained from  a Heisenberg
Hamiltonian  by  application of  a  mean  field  approach for  a  spin
cluster, as  well as from  a many-body Green's function  theory within
the Tyablikov-decoupling (RPA).
\end{abstract} 
 \maketitle
                                      
\section{Introduction}
The  investigation   of  the  magnetic   properties  of  ferromagnetic
ultrathin films  is a field of intense  current interest \cite{HeB94}.
In this contribution we  determine the magnetic in-plane anisotropy of
a square  two-dimensional (2D)  Heisenberg ferromagnet in  presence of
the magnetic dipole interaction.  In 2D magnets the action of magnetic
anisotropies is twofold. First, they induce a long range magnetization
\cite{Kit51,Mal76,LHR99} with a Curie  temperature of the order of the
exchange  coupling  \cite{Yab91},  whereas  a  strictly  isotropic  2D
Heisenberg  magnet does  not exhibit any long range magnetic  order at
finite  temperatures (Mermin-Wagner-theorem)  \cite{Mer66}. Moreover,
the long range character of the dipole interaction itself is
sufficient to stabilize the magnetization in a 2D magnet. Secondly,
the anisotropies determine the easy and hard axes of the magnetization
with respect  to the lattice frame \cite{HeB94}.   For a ferromagnetic
thin film  the dipole  coupling prefers an  in-plane direction  of the
magnetization  (shape anisotropy,  demagnetizing  field).  Except  for
very  small thicknesses  thin films  are usually  in-plane magnetized.
The corresponding  in-plane easy axis depends on  the lattice symmetry
of  the film  face.   For  a rectangular  (110)  monolayer the  dipole
interaction  induces a  uniaxial  in-plane anisotropy,  the easy  axis
being the in-plane axial  direction with the smaller lattice constant,
as can easily be shown by calculating the corresponding lattice sums.

On the other hand, for a \textit{square} monolayer the dipole coupling,
although itself  not  rotational invariant, exhibits  a 
\textit{continuous energy degeneracy}  for classical spins at  $T=0$. 
Thus,  due  to  the  Mermin-Wagner theorem  a  long  range
magnetic order is not expected at finite temperatures. However, as has
been  shown by  Monte  Carlo calculations  and  interacting spin  wave
theory,  a magnetic  ordering and  a critical  temperature  exists for
dipole  coupled spins  on  a square  lattice  \cite{DMB97}, since  the
\textit{magnetic excitations} are not continuously degenerate. In this
case a quartic in-plane  anisotropy is present, the corresponding easy
axes  being the  edges  of the  square  lattice. In  other words,  the
density  of  states and  thus  the  entropy  depends on  the  magnetic
direction within  the lattice.  This phenomenon is  an example  of the
\textit{'order-by-disorder'}    effect     in    frustrated    magnets
\cite{Vil79}.   In  these  systems  thermal  fluctuations  or  lattice
disturbances  partly remove  frustrations, and  a  collective magnetic
ordering emerges at finite temperatures.

Whereas  the  existence  of  a  finite  magnetization  and  a  quartic
anisotropy at finite temperatures for dipole coupled spins on a square
lattice has been proven, the  strength and the temperature behavior of
this anisotropy has not been determined yet. Of particular interest is
whether   the  dipole-coupling-induced   quartic  anisotropy   can  be
measured, and how this anisotropy compares with the single-ion quartic
lattice anisotropy due  to the spin-orbit coupling.  In  this study we
will  adress   these  questions  with   the  help  of  a   mean  field
approximation,  taking  into  account   a  square  spin  cluster  with
different  numbers  of  spins (Oguchi-theory)  \cite{Ogu55,Sma66},  as
outlined  in  Sec.II.   Furthermore,  we  apply  a  many-body  Green's
function theory  within the Tyablikov-decoupling  (RPA) by considering
collective magnetic excitations  (spin waves) and interactions between
them  \cite{Tya59,Tya67}, which  are  in particular  important for  2D
magnetic systems. Our results are summarized and discussed in Sec.III.

\section{Theory}
The  free energy  $F(T,\phi)$ is  determined within  a Heisenberg-type
Hamiltonian  as a  function of  the temperature  $T$ and  the in-plane
angle $\phi$.   A square (001) ferromagnetic  monolayer is considered,
which  is spanned by  the $xz$-  plane, the  $z$- direction  refers to
$\phi=0$. The isotropic  nearest neighbor Heisenberg exchange coupling
$J$, the magnetic dipole-dipole  interaction, and an external magnetic
field \textbf{B} between  localized Heisenberg spins $\mathbf{S}_i$ on
lattice sites $i$ are taken into account:
\begin{eqnarray}
\mathcal{H}&=&-\frac{J}{2}\;\sum_{\la i,j\ra}\mathbf{S}_i\,\mathbf{S}_j
-g\mu_B\sum_i\mathbf{B}\;\mathbf{S}_i \nonumber \\ 
&& \hspace*{-1.5cm} +\frac{(g\mu_B)^2}{2} 
\sum_{i,j\atop i\ne j} \frac{1}{r_{ij}^5}\Big[r_{ij}^2\,
\mathbf{S}_i\,\mathbf{S}_j-3\,\Big(\mathbf{r}_{ij}\,\mathbf{S}_i\Big)
\Big(\mathbf{r}_{ij}\,\mathbf{S}_j\Big)\Big] \;, \label{e1}
\end{eqnarray}
with $g$  the Land\'e factor and  $\mu_B$ the Bohr  magneton. The spin
quantum number is  assumed to be $S=1/2$.  The  lattice vector between
sites $i$ and $j$ is given by $\mathbf{r}_{ij}$, with
$r_{ij}=|\mathbf{r}_{ij}|$ the  distance. All spins are  assumed to be
aligned    parallely   (ferromagnetic   order),    the   magnetization
$\mathbf{m}_i(T)=\mathbf{m}(T)=(0,0,m(T))$   is  directed   along  the
magnetic field $\mathbf{B}=(0,0,B)$.

To account  for a varying angle  of the magnetization  with respect to
the lattice frame, the lattice is rotated by the in-plane angle $\phi$
with respect to the $z$- axis, yielding the rotated lattice vectors:
\begin{equation}
\mathbf{r}'_{ij}=\left(\begin{array}{c}r_{ij}^z\,\sin\phi+
r_{ij}^x\,\cos\phi \\ 0 \\ 
r_{ij}^z\,\cos\phi-r_{ij}^x\,\sin\phi\end{array} \right) \,. 
\label{e1a} \end{equation}
Clearly,  the only  source in  Eq.(\ref{e1}) for  a  possible in-plane
anisotropy  is the double  scalar product  of the  dipole interaction.
Within  the  single-spin  (Bragg-Williams-) mean  field  approximation
\cite{Sma66} the dipole coupling  yields an additional contribution to
the molecular field. The resulting single-spin Hamiltonian reads

\begin{eqnarray}
\mathcal{H}^{(1)}&=&-\Big(q\,J\,m(T)+w\,S(0,0)\,m(T)+g\mu_B\,B\Big)\,
S_1^z \nonumber \\ && 
+\frac{1}{2}\,m^2(T)\,\Big(q\,J+w\,S(0,0)\Big) \;, \label{e2}
\end{eqnarray}
with  $q=4$ the  number  of nearest  neighbors (coordination  number),
$w=(g\mu_B)^2/a_0^3$ the strength of the dipole interaction, $a_o$ the
lattice constant, $S(0,0)\sim\, 4.51681$ the corresponding lattice sum
\cite{BeM69}, and $m(T)=\la  S_1^z\ra$ the magnetization. Thus, within
this approximation the dipole  interaction is \textit{isotropic} for a
square monolayer and does not depend on the in-plane angle $\phi$.

Our approach  will now  be improved by  applying the  so-called Oguchi
method  \cite{Ogu55,Sma66}. A  number  $N>1$ of  neighboring spins  is
considered, the interactions between the $N$ spins in this cluster are
treated exactly. The remaining lattice  is coupled to the spin cluster
by a molecular field.  To conserve the symmetry of the square lattice,
we  take into  account  only square-shaped  spin  clusters, the  three
smallest  possible clusters  are characterized  by $N=4,\;9$,  and 16.
Note that  the resulting cluster  Hamiltonians $\mathcal{H}^{(N)}$ are
usually non-diagonal,  i.e.\ contributions proportional  to the $S^x$-
and $S^y$-  spin components may  be present.  Therefore, as  a further
approximation we  consider only  the diagonal elements.  This approach
guarantees  that  the magnetization  is  directed  along the  external
magnetic  field. As  an  example, the  Hamiltonian  for the  four-spin
cluster ($N=4$) reads:
\begin{widetext} \begin{eqnarray} \mathcal{H}^{(4)}
&=&-(J-w)\Big(S_1^zS_2^z+S_1^zS_3^z+S_2^zS_4^z+S_3^zS_4^z\Big)
-\frac{w}{2^{5/2}}(S_1^zS_4^z+S_2^zS_3^z) \nonumber \\ 
&& -3\,w\,\bigg[\sin^2\phi\,(S_1^zS_2^z+S_3^zS_4^z)
+\cos^2\phi\,(S_1^zS_3^z+S_2^zS_4^z)
+\frac{2\,\cos\phi\sin\phi}{2^{5/2}}(S_1^zS_4^z-S_2^zS_3^z)\bigg]
\nonumber \\
&&-\Big(2\,J\,m(T)+g\mu_B\,B\Big)\Big(S_1^z+S_2^z+S_3^z+S_4^z\Big)
-w\,m(T)\,(S_1^z+S_4^z)\Big(S(0,0)-1-
\frac{1}{2^{5/2}}(1+6\,\cos\phi\,\sin\phi)\Big) \nonumber \\ 
&&\hspace*{-1cm} -w\,m(T)\,(S_2^z+S_3^z)\Big(S(0,0)-1-
\frac{1}{2^{5/2}}(1-6\,\cos\phi\,\sin\phi)\Big) 
+4\,J\,m^2(T)+2\,w\,m^2(T)\,\Big(S(0,0)-1-
\frac{1}{2^{5/2}}\Big) \,. \label{e11} \end{eqnarray}
\end{widetext}
The partition function for this case is given by:
\begin{equation}
Z^{(4)}(T,B,\phi)=\sum_{S_1^z,S_2^z,S_3^z,S_4^z=-S}^{+S}
\;\exp(-\beta\mathcal{H}^{(4)}) \,, \label{e12} \end{equation}
and the (average) magnetization by:
\begin{eqnarray}
m(T,B,\phi)&=&\frac{1}{4\,Z^{(4)}(T,B,\phi)} \times \nonumber \\ 
&& \hspace*{-2.7cm} 
\sum_{S_1^z,S_2^z,S_3^z,S_4^z=-S}^{+S}\;(S_1^z+S_2^z+S_3^z+S_4^z)\;
\exp(-\beta\mathcal{H}^{(4)}) \,, \label{e13} \end{eqnarray}
with   $\beta=1/k_BT$   and  $k_B$   the   Boltzmann  constant.    The
corresponding expressions for the  other investigated sizes are rather
lengthy  and  not shown  here.   The  difference  of the  free  energy
$F(T,B,\phi)=-k_BT\,\ln    Z(T,B,\phi)$     between    the    diagonal
($\phi=\pi/4$) and the axial directions ($\phi=0$) yields the in-plane
anisotropy $\mathcal{K}_{4,dip}(T,B)$,  which will be  calculated as a
function of the temperature $T$ and the external magnetic field $B$.

It is important to mention that  the mean field theory as described in
the preceeding  subsection does not  fulfill the Mermin-Wagner-theorem
for the isotropic  2D Heisenberg magnet.  In this  case the long range
magnetic   order   becomes   unstable  against   collective   magnetic
excitations with long wavelengths \cite{Kit51,Mal76,Yab91}.  Thus, the
consideration of these  spin waves is very important  for the magnetic
behavior of  a ferromagnetic monolayer.  For the  determination of the
dipole-coupling-induced  in-plane anisotropy  we apply  in  addition a
many-body  Green's function theory  to the  Hamiltonian Eq.(\ref{e1}),
after  the   lattice  is  rotated   by  the  in-plane   angle  $\phi$,
Eq.(\ref{e1a}).      We     consider     the     Green's     functions
$G_{ij}^{\alpha-}=\la\la    S_i^\alpha;S_j^-\ra\ra$,   $\alpha=+,-,z$,
which are  solved within the  Tyablikov-decoupling \cite{Tya59}. Since
interactions  between spin waves  are partly  taken into  account, the
magnetic properties can  be determined up to the  Curie temperature by
this  method.  The respective  formalism  is  described  in detail  in
\cite{LHR99},  thus  we outline  merely  necessary  extensions in  the
Appendix.   The  magnetization   $m(T,B,\phi)$,  the  internal  energy
$E(T,B,\phi)$, and the free  energy $F(T,B,\phi)$ are calculated.  For
comparison,  we   determine  the  corresponding   quantities  also  by
considering  the Holstein-Primakoff approximation  \cite{HoP40}, which
is valid at low temperatures.

\section{Results and Discussion}
By  application of the  described methods  we calculate  the effective
dipole-coupling-induced       in-plane       magnetic       anisotropy
$\mathcal{K}_{4,dip}(T,B)$ for  a square ferromagnetic  monolayer as a
function of the temperature $T$ and the applied magnetic field $B$.  A
finite value indicates that this  anisotropy is caused by the magnetic
dipole coupling, since other  sources of magnetic anisotropies are not
present  in the Hamiltonian  Eq.(\ref{e1}).  For  the strength  of the
dipole  interaction we  choose $(g\mu_B)^2/a_o^3=w=0.01\,J$,  which is
appropriate         for         $3d$-         transition         metal
ferromagnets.  $\mathcal{K}_{4,dip}(T,B)$  is given  in  units of  the
energy   difference  between  the   perpendicular  and   the  in-plane
magnetization                  (demagnetizing                  energy)
$E_{demag}(0)=(3/2)\,w\,S(0,0)\,m^2(0)\approx   2\pi\,w\,S^2$   for  a
ferromagnetic monolayer at $T=0$.
 
The free energy $F(T,\phi)$ exhibits a four-fold symmetry 
as a  function of  the in-plane  angle $\phi$.   The easy
magnetic axes are directed along the edges ($\phi=0,\pi/2$, etc.), and
the hard axes along  the diagonals ($\phi=\pi/4,3\pi/4$, etc.)  of the
square  lattice as  obtained in  \cite{DMB97}. The  quartic anisotropy
$\mathcal{K}_{4,dip}(T,B)$ is  depicted in Fig.1 as a  function of the
temperature, calculated by the Oguchi approach.  Three different sizes
of  the  spin cluster  are  considered  ($N=4,\;9,\;16$),  as well  as
different magnetic field strengths $B$. The Curie temperature $T_C(N)$
decreases with  an increasing $N$, since  additional spin correlations
are   considered.    As   can  be   seen,   $\mathcal{K}_{4,dip}(T,B)$
\textit{increases} with an  increasing temperature, reaching a maximum
at $T_C$.  This temperature  behavior is in  striking contrast  to the
corresponding  behavior of  other anisotropic  contributions.  Usually
the  effective  anisotropies   \textit{decrease}  with  an  increasing
temperature,      and     vanish      at     $T_C$      for     $B=0$.
$\mathcal{K}_{4,dip}(T,B)$ exhibits a cusp  at $T_C$ and decreases for
temperatures $T>T_C$.  By increasing the size $N$ of the spin cluster,
$\mathcal{K}_{4,dip}(T,B)$  becomes  larger.    An  application  of  a
magnetic field  reduces $\mathcal{K}_{4,dip}(T,B)$, and  the cusp near
$T_C$  changes  to  a  maximum.  For comparison,  in  Fig.1  we  show
$\mathcal{K}_{4,dip}(T,B)$  for $B=0$  and $N=4$  by  considering also
non-diagonal elements in  the cluster Hamiltonian $\mathcal{H}^{(4)}$,
cf.\   Sec.II.   A   decrease    of   $T_C$   and   an   increase   of
$\mathcal{K}_{4,dip}(T,B)$ by  $\sim10\%$ is obtained  with respect to
the calculations which consider diagonal elements only.
\begin{figure} \label{fig1}
\includegraphics[width=8cm]{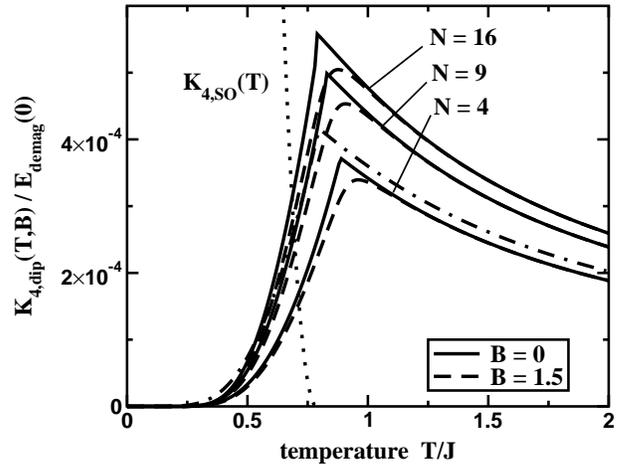}
\caption{
Effective dipole-coupling-induced quartic in-plane magnetic anisotropy
$\mathcal{K}_{4,dip}(T,B)$ for a square monolayer as a function of the
temperature $T$,  calculated within the  Oguchi mean field  method for
spin  cluster  sizes  $N=4,\;9,\;16$.   Two different  magnetic  field
strengths  $B$ are  considered  ($g\mu_BB/E_{demag}(0)=0,\;1.5$).  The
field energy and $\mathcal{K}_{4,dip}(T,B)$  are given in units of the
demagnetizing  energy  $E_{demag}(0)$.    The  dot-dashed  line  shows
$\mathcal{K}_{4,dip}(T,B)$   for  $B=0$   and  $N=4$   by  considering
non-diagonal  matrix elements  of the  Hamiltonian, cf.\  Sec.II.  For
comparison,   also  the   effective   single-ion  quartic   anisotropy
$\mathcal{K}_{4,so}(T)$ resulting  from the spin-orbit  interaction is
shown, calculated  within a thermodynamic perturbation  theory by 
assuming $\mathcal{K}_{4,so}(0)=0.01\,E_{demag}(0)$ (dotted line). 
} \end{figure}

In   addition,  we   have  calculated   $\mathcal{K}_{4,dip}(T,B)$  by
application of  the many-body Green's  function theory as  outlined in
\cite{LHR99}  and   in  the  Appendix.   For   $w=0.01\,J$  the  Curie
temperature is calculated to be $k_BT_C^{RPA}=0.373\,J$. In Fig.2(a,b)
we  present  the   resulting  magnetization  $m(T,B)$  and  anisotropy
$\mathcal{K}_{4,dip}(T,B)$  as functions  of the  temperature  and the
magnetic  field.  In  accordance with  the results  obtained  from the
Oguchi   approach,   $\mathcal{K}_{4,dip}(T,B)$   increases  with   an
increasing   temperature.   Due  to   the  consideration   of  quantum
mechanical  spins  a  finite  value of  $\mathcal{K}_{4,dip}(T,B)$  is
already present  at $T=0$, resulting  from the zero-point  spin motion
(quantum fluctuations).  Note that the presence of the dipole coupling
causes      a      non-saturated     magnetization      $m(T=0)<S=1/2$
\cite{Mal76,HoP40}.  In  Fig.2 also the  results as obtained  from the
Holstein-Primakoff    approximation   are   shown,    which   neglects
interactions  between  spin  waves.  The  corresponding  magnetization
$m(T,B)$ and anisotropy $\mathcal{K}_{4,dip}(T,B)$ at low temperatures
are  close  to the  results  as obtained  from  the  RPA. At  elevated
temperatures $T\gtrsim T_C/3$  the Holstein-Primakoff approximation is
no longer valid.
\begin{figure} \label{fig2} 
\includegraphics[width=8cm]{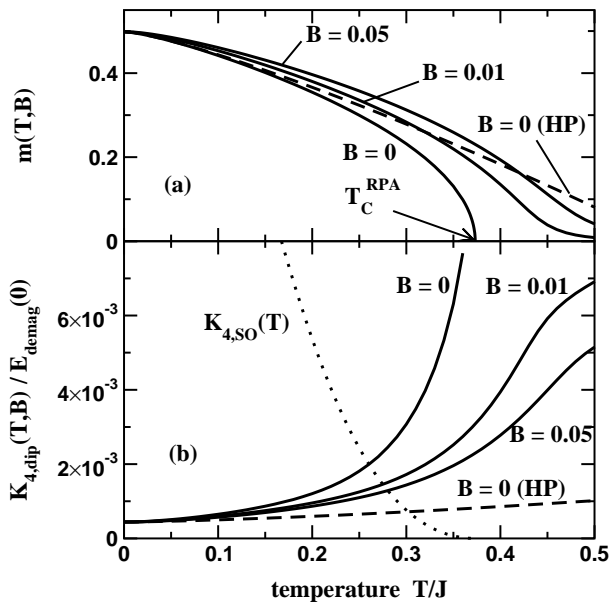}
\caption{
Magnetization  $m(T,B)$ (a)  and effective  quartic  in-plane magnetic
anisotropy  $\mathcal{K}_{4,dip}(T,B)$ (b) for  a square  monolayer as
functions  of  the  temperature   $T$  and  the  magnetic  field  $B$,
calculated by  application of the  Green's function method  within the
Tyablikov-decoupling (RPA). The Curie  temperature is calculated to be
$k_BT_C^{RPA}=0.373\,J$.        The       field       energy       and
$\mathcal{K}_{4,dip}(T,B)$  are given  in units  of  the demagnetizing
energy $E_{demag}(0)$.  In addition,  the results as obtained from the
Holstein-Primakoff approximation  (HP) for $B=0$  are depicted (dashed
lines). For comparison, the dotted line shows the effective single-ion
quartic anisotropy   $\mathcal{K}_{4,so}(T)$, with 
$\mathcal{K}_{4,so}(0)=0.01\,E_{demag}(0)$.  } \end{figure}

As  has  already  been  obtained  from Monte  Carlo  calculations  and
interacting spin wave theory \cite{DMB97}, the dipole-coupling-induced
in-plane   anisotropy  is  caused   by  magnetic   fluctuations.  This
order-by-disorder effect is thus controlled by the magnetic entropy of
the system \cite{Vil79}. An energetic influence should become apparent
at  $T=0$ already  within  a mean  field  approximation.  Besides  the
effects resulting from the zero-point  motion, the terms of the dipole
interaction  dependent  on the  in-plane  angle  $\phi$  cancel for  a
strictly  square  magnetic  lattice,   i.e.\  for  a  fully  saturated
ferromagnetic  state.  At  finite  temperatures  thermal  fluctuations
disturb this square periodicity, causing an effective quartic in-plane
magnetic  anisotropy, which  reflects the  underlying  spatial lattice
symmetry.  The single-spin mean field theory conserves the periodicity
of the  square magnetic  lattice, since all  other spin  operators are
replaced   by   uniform   expectation   values.  Thus,   within   this
approximation   one  obtains   $\mathcal{K}_{4,dip}(T,B)=0$   for  all
temperatures.   On the  other  hand, within  the  Oguchi approach  the
interactions  between   a  number   of  spins  are   treated  exactly,
considering   a  few  collective   magnetic  excitations   with  short
wavelengths ranging over several lattice constants, and resulting in a
finite value  for $\mathcal{K}_{4,dip}(T,B)$ for  $T>0$. The many-body
Green's  function  theory  takes  into  account spin  waves  with  all
possible wavelengths.  Since  $\mathcal{K}_{4,dip}(T,B)$ is maximal at
elevated   temperatures,  $\mathcal{K}_{4,dip}(T,B)$   is   caused  by
magnetic  fluctuations  in particular  with  short wavelengths,  which
become excited in this temperature range.

Within  the   Oguchi  approach  we   calculate  a  maximum   value  of
$\mathcal{K}_{4,dip}(T,B)$ of the order  of 0.1\% of the demagnetizing
energy $E_{demag}(0)$.  On the other  hand, $\mathcal{K}_{4,dip}(T,B)$
as  calculated  from  the  Green's  function method  is  more  than  a
magnitude                        larger,                        namely
$\mathcal{K}_{4,dip}(T,B)\,\sim\,0.01\,E_{demag}(0)$    near    $T_C$.
Similarly,  the action  of  a  magnetic field  $B$  on the  anisotropy
$\mathcal{K}_{4,dip}(T,B)$  is calculated to  be much  more pronounced
within the  Green's function theory  than within the  Oguchi approach,
cf.\ Figs.1 and 2.  The  reason for the strong differences between the
results obtained from the two methods is that the former one considers
collective  magnetic  excitations with  long  wavelengths.  These  are
known to have strong  effects on the magnetic properties of ultrathin
films \cite{Mal76,Yab91}. In this case the magnetic field acts merely 
on \textit{spin blocks}, i.e.\ correlated  regions of  neighboring  
spins characterized  by the short range order  parameter \cite{Pol75}, 
resulting in a much larger magnetic response. 
For instance,  the induced  magnetization in a  ferromagnetic trilayer
has  been   determined  to  be   an  order  of  magnitude   larger  by
consideration  of spin  waves \cite{JBP99}.   The  collective magnetic
excitations  are   most  pronounced   for  a  single   magnetic  layer
\cite{Mal76}, the  zero-point motion is strongest for  the spin number
$S=1/2$.  Note that  the theoretical  methods applied  in  the present
study   are  much   less  demanding   than  Monte   Carlo  simulations
\cite{DMB97}.

The obtained  dipole-coupling-induced in-plane anisotropy  is small as
compared to  other (effective) anisotropies, since  we have considered
interaction   strengths  appropriate   for   $3d$-  transition   metal
ferromagnets. A   small  value  for   $\mathcal{K}_{4,dip}(T,B)$  has  
been conjectured in \cite{DMB97}. 
If  a ten  times larger ratio  $w/J$ between  the dipole
coupling strength  and the exchange interaction is  assumed, the ratio
$\mathcal{K}_{4,dip}(T,B)/E_{demag}(0)$ increases  roughly by the same
factor.   Since usually the effective anisotropies decrease  with  an
increasing   temperature,   $\mathcal{K}_{4,dip}(T,B)$  might   become
observable at elevated temperatures. For comparison, we show in Figs.1
and  2  also  the  effective single-ion  quartic  in-plane  anisotropy
$\mathcal{K}_{4,so}(T)$  resulting  from  the spin-orbit  interaction.
This  quantity  is  calculated   with  the  help  of  a  thermodynamic
perturbation  theory  \cite{Cal66}. By  assuming  its  strength to  be
$\mathcal{K}_{4,so}(0)=0.01\,E_{demag}(0)$, and adapting corresponding
Curie  temperatures,  $\mathcal{K}_{4,so}(T)$  becomes  comparable  to
$\mathcal{K}_{4,dip}(T,B)$  at $T/T_C\sim 0.8  - 0.9$. In this
temperature range the \textit{total} quartic in-plane anisotropy
should increase again, or exhibit an in-plane magnetic reorientation. 
As mentioned, this  dipole-coupling-induced  anisotropy should be 
more apparent for ferromagnetic (001) thin films with a large $w/J$- 
ratio. 

We like  to comment on the  fact that a finite  anisotropy is obtained
for  $T>T_C$.  Usually  the  effective anisotropy  as  observed for  a
collectively  ordered   ferromagnetic  state  vanishes   above  $T_C$.
However,  a  vanishing  effective  anisotropy  for  $T>T_C$  does  not
indicate that the anisotropy  as present in the Heisenberg Hamiltonian
disappears. A single (paramagnetic) spin is still subject to e.g.\ the
single-ion uniaxial anisotropy $K_2$  also if the net magnetization is
zero. The  resulting free energy  difference between the easy  and the
hard  magnetic  directions   ('paramagnetic  anisotropy')  behaves  as
$\propto K_2^2/k_BT$ for  $K_2\ll k_BT$, and is rather  small if $K_2$
is   small  as   compared  to   the  exchange   interaction  $J\propto
k_BT_C$. With regard  to the present study, a finite value of    
$\mathcal{K}_{4,dip}(T,B=0)$    for    $T>T_C$   reflects the
dipole-coupling-induced anisotropy of a spin block \cite{Pol75}. 

Note  that the  free  energy  $F(T,B)$ as  obtained  from the  Green's
function method  yields unphysical  results for temperatures  near and
above $T_C$.  As can be shown \cite{Jenup}, for large temperatures the
free energy  $F(T,B)$ as calculated  by this method does  not approach
the  value  $-k_BT\,\ln2$, which  is  the  free  energy of  a  single,
non-interacting $S=1/2$- spin.  This  unphysical behavior results in a
still   increasing   $\mathcal{K}_{4,dip}(T,B)$   for  $T>T_C$,   cf.\
Fig.2(b).  Nevertheless, we expect  that the Green's function approach
reflects  the  correct   behavior  of  $\mathcal{K}_{4,dip}(T,B)$  for
$T<T_C$, since its  behavior is corroborated by the  Oguchi method and
the Holstein-Primakoff approach.

We have considered a square monolayer only. A similar effect is expected
also for  thicker films with a  square (001) face, or  for a hexagonal
(111) thin film. In addition, a  corresponding effect may emerge for a
three-dimensional  cubic  lattice, for  which  the dipole  interaction
cancels exactly.  At elevated temperatures, however, caused by thermal
fluctuations  the   frustration  due  to  this   periodicity  will  be
lifted. The  dipole coupling then induces a  cubic magnetic anisotropy
with  easy  axes   directed  along  e.g.\  the  edges   of  the  cubic
lattice. Also is this case an increasing anisotropy with an increasing
temperature is expected. To  obtain an anisotropic contribution caused
by  the magnetic  dipole interaction via the order-by-disorder effect,  
we emphasize  that  the lattice
symmetry must be strictly cubic or square.  A lattice distortion e.g.\
due to magnetostrictive effects will probably blur this effect.

P.J.J.\ gratefully acknowledges the  invitation and the hospitality of
the I.\  Institute for Theoretical Physics of  the Hamburg University.
Fruitful  discussions with  Prof.\ K.D.\  Schotte, Free  University of
Berlin, are acknowledged.

\section*{Appendix: Many-body Green's function theory}
The following commutator Green's  functions in the frequency space are
taken        into        account:        $G_{ij}^\alpha(\omega)=\la\la
S_i^\alpha;S_j^-\ra\ra_\omega$,  $\alpha=+,-,z$.   By considering  the
Hamiltonian Eq.(\ref{e1}),  these Green's functions are  solved in the
usual  way by  the equation  of motion  \cite{LHR99,Tya59,Tya67}.  The
higher   order    Green's   functions   are    approximated   by   the
Tyablikov-decoupling     (RPA)     \cite{Tya59},     e.g.\     $\la\la
S_k^z\,S_i^+;S_j^-\ra\ra\simeq\la  S^z\ra \la\la  S_i^+;S_j^- \ra\ra$,
resulting    in   a    system    of   linear    equations   for    the
$G_{ij}^\alpha(\omega)$.  By application  of a  Fourier transformation
into  the 2D  wave vector  space  $\mathbf{k}_\|\equiv\mathbf{k}$, one
obtains $G^z(\mathbf{k},\omega)=0$, if $\la S^+\ra=\la S^-\ra=0$.  The
remaining two Green's  functions $G^\pm(\mathbf{k},\omega)$ are easily
evaluated \cite{LHR99}:
\begin{eqnarray} 
G^+(\mathbf{k},\omega) &=& 2\,m(T)\;\frac{\omega+a(\mathbf{k})}
{\omega^2-\varepsilon^2(\mathbf{k})} \,, \label{a2} \\
G^-(\mathbf{k},\omega) &=& 2\,m(T)\;\frac{b(\mathbf{k})}
{\omega^2-\varepsilon^2(\mathbf{k})} \,, \label{a3} \end{eqnarray}
with    $m(T)=\la   S^z\ra$.    The    magnon   dispersion    relation
$\varepsilon(\mathbf{k})$ is given by
\begin{eqnarray} 
\varepsilon^2(\mathbf{k})&=&a^2(\mathbf{k})-b^2(\mathbf{k})
\,,\label{a4} \\ a(\mathbf{k}) &=& 
g\mu_BB+m(T)\Big(J(0)-J(\mathbf{k})\Big) \label{a5} \\ 
&& \hspace*{-1.2cm} +\,w\,m(T) \bigg[S(0,0)+S(k_z,k_x)\,\Big(1-\frac{3}{2}
\cos^2\phi\Big) \nonumber \\ && \hspace*{-1.2cm} 
+\,S(k_x,k_z)\,\Big(1-\frac{3}{2}\sin^2\phi\Big)+3\,T(k_z,k_x)\,
\cos\phi\,\sin\phi\bigg] \,, \nonumber \\
b(\mathbf{k})&=&\frac{3}{2}\,w\,m(T)\Big[
S(k_z,k_x)\,\cos^2\phi+S(k_x,k_z)\,\sin^2\phi \nonumber \\ &-& 
2\,T(k_z,k_x)\,\cos\phi\,\sin\phi \Big] \,. \label{a6} \end{eqnarray}
$J(\mathbf{k})=2\,J\,(\cos   k_xa_0+\cos  k_za_0)$   is   the  Fourier
transform  of  the  exchange   interaction,  with  $a_o$  the  lattice
constant. The oscillating lattice sums \cite{BeM69} are defined by
\begin{eqnarray} 
S(k_x,k_z) &=& \sum_{l,n=-\infty}^{+\infty}
\frac{l^2}{(l^2+n^2)^{5/2}}\;\exp(-ik_xa_0l-ik_za_0n) \,, \nonumber \\
T(k_z,k_x) &=& \sum_{l,n=-\infty}^{+\infty}
\frac{l\,n}{(l^2+n^2)^{5/2}}\;\exp(-ik_za_0l-ik_xa_0n)\,, \nonumber \\
\label{a8} \end{eqnarray} 
where  the terms  with  $l=n=0$  have to  be  omitted.  The  following
expressions  are  valid for  the  spin  quantum  number $S=1/2$.   The
magnetization   $m(T)$   is  obtained   from   the  spectral   theorem
\cite{Tya67}:
\begin{eqnarray} 
m(T,\phi) &=& \frac{1}{2}-\frac{1}{N} \sum_\mathbf{k} 
\la S^-S^+\ra(\mathbf{k}) \nonumber \\ 
&=& \frac{1}{2}-\frac{m(T)}{N} \sum_\mathbf{k}
\Big(\frac{a(\mathbf{k})}{\varepsilon(\mathbf{k})}\,\coth x-1\Big) \,,
\label{a11} \end{eqnarray} 
with  $x=\varepsilon(\mathbf{k})/2k_BT$.  The  summation  extends over
the   first    Brillouin   zone,    $N$   denotes   the    number   of
$\mathbf{k}$-points.   By  calculating   the  expectation  value  $\la
S_i^zS_j^z\ra$ with the help of $G^+(\mathbf{k},\omega)$ \cite{Tya67},
we obtain for the internal energy $E(T,\phi)$ per spin:
\begin{eqnarray} 
E(T,\phi) &=& E_0+\frac{m(T,\phi)}{2\,N} \sum_\mathbf{k}\Bigg(
e_1(\mathbf{k})\Big(\frac{a(\mathbf{k})}{\varepsilon(\mathbf{k})}
\,\coth x-1\Big) \nonumber \\ 
&& \hspace*{-1.5cm}
+\Big(e_2(\mathbf{k})\,\frac{b(\mathbf{k})} {\varepsilon(\mathbf{k})}
+\varepsilon(\mathbf{k})\Big)\coth    x   -a(\mathbf{k})+b(\mathbf{k})
\Bigg) \,, \label{a12} \end{eqnarray}
with the denotations
\begin{eqnarray} 
E_0 &=& -\frac{1}{2} g\mu_BB-\frac{1}{8}\Big(J(0)+w\,S(0,0)\Big)
\,,\label{a13} \\
e_1(\mathbf{k}) &=& g\mu_BB+\frac{1}{2}\Big(J(0)-J(\mathbf{k})\Big)
\nonumber \\ && +
\frac{w}{2}\,\Big(S(0,0)+S(k_z,k_x)+S(k_x,k_z)\Big)\,,\label{a14} \\
e_2(\mathbf{k}) &=& \frac{1}{2}\,J(\mathbf{k})- \frac{w}{2}\,
\Big(S(k_z,k_x)+S(k_x,k_z) \Big) \,. \label{a15}
\end{eqnarray} 
Finally, the free energy  $F(T,\phi)$ is calculated from a temperature
integral   over  $E(T,\phi)$  \cite{Tya67}.    The  Holstein-Primakoff
approximation \cite{HoP40} is  obtained by replacing the magnetization
$m(T)$ on  the right  side of Eq.(\ref{a11})  by its  saturation value
$S=1/2$.

\end{document}